# The Regenerative Current Mirror: A Very Low Power Front-End Amplifier for Silicon Pixel Detectors

Jinyuan Wu

*Abstract*— **In pixelized detectors, reducing power consumption in the front end ASIC chips becomes a crucial demand. Optimization based on mature pre-amplifier schemes today is unlikely to bring sufficient improvements. A new CMOS front-end gain stage topology with very low power consumption called regenerative current mirror is developed to fulfill the demand. The circuit takes advantage of high speed performance of current amplification while operating with relatively low bias current. The regenerative current mirror uses a NMOS current mirror and a PMOS current mirror, both with nominal gain of 1, to form a loop-back topology that provides a positive feedback. An NMOS FET with an external adjustable voltage applied to it gate terminal is used to limit open loop gain of the current mirrors to be slightly lower than 1. This yields a net gain of the positive feedback loop to be much larger than 1 while operating the current mirrors under relatively low bias currents. Simulation shows that in 65 nm fabrication process, the power consumption of the gain stages suitable for silicon pixel detectors can be controlled < 10 micro-Watts per channel.**

## I. Introduction

In pixelized detector used in high energy physics experiments, power consumption in the front-end ASIC becomes a critical issue. Since the number of detector elements becomes so large, very low power consumption of the front-end circuits is required. In today's mature pre-amplifier scheme, the power consumption is around 800 micro-Watts per channel in 65 nm CMOS fabrication technology, while the desired power consumption in future ASIC for pixelized detectors can be as low as 10 micro-Watts.

In typical small signal voltage amplifiers, transistors must operate with a sufficiently high bias current so that the transistors are in high conductivity region to follow the input pulse at high speed. Also, the topologies used in amplifiers usually employs deep negative feedback to improve bandwidth and linearity. Given these basic principle, it is unlikely to achieve demanded improvements by optimizing on the existing topologies.

As an attempt of significantly reducing power consumption of the front-end gain stage, a new CMOS front-end topology with very low power consumption called regenerative current mirror is developed. The simplified schematics of the circuit is shown in Fig. 1.

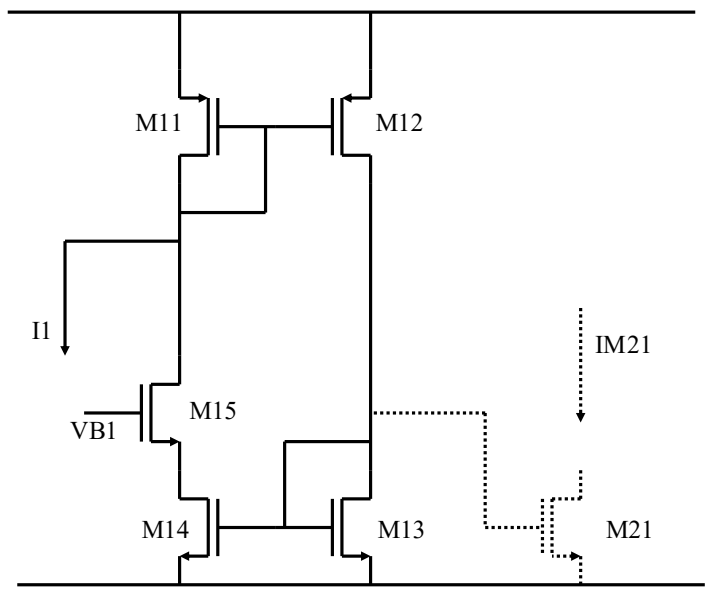

Fig. 1. The regenerative current mirror

In this circuit, MOS FET pairs (M11, M22) and (M13, M14) form two current mirrors. The two current mirrors are cross connected to form a positive feedback loop. The transistors in each current mirror are identical so that the nominal gain of each current mirror is 1.

Obviously, should the total gain of the cross connected current mirrors be larger than 1, the circuit would be locked to the short circuit condition in which all the FETs would be conductive. To prevent the locking condition from happening, the NMOS FET M15 is used to limit the gain of the NMOS current mirror so that it is <1.

With the external voltage VB1 adjusted appropriately, the total open loop gain A of the two current mirrors can be set to be slightly <1. In this case, the close loop gain of the whole circuit G = A/(1-A). If A=0.8, for example, G = 4.

The current, primarily bias current, flowing through M11, M15 and M14 path is the same as the M12 and M13 path, rather than G times higher in regular current mirror structure. In this scheme, the bias currents in both paths remain unchanged and hence reduces power consumption.

## II. Simulation of the Gain Stage

The actual circuit diagram of a single stage of the regenerative current mirror simulated with the Spice software is shown in Fig. 2. The 65 ns CMOS device model is used in The author is with Fermi National Accelerator Laboratory, Batavia, IL 60510 USA (phone: 630-840-8911; fax: 630-840-2950; e-mail: jywu168@ fnal.gov).

the simulation. To conveniently study the power consumption of the circuit, a small internal resistance is introduced in the power supply V1 and a bypass capacitor C2 is also included.

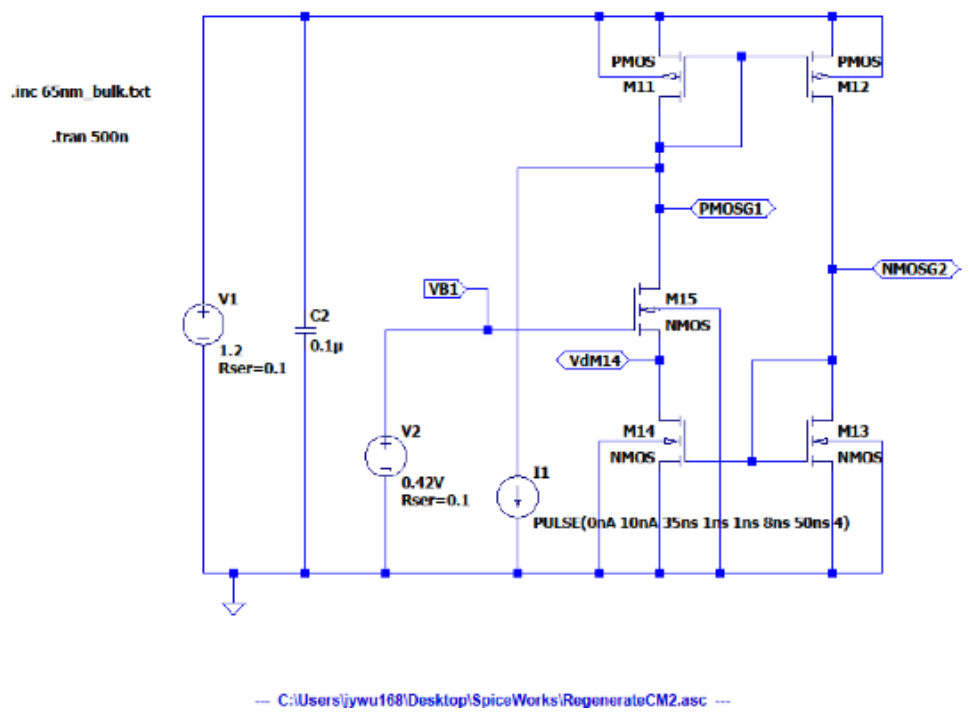

Fig. 2. Circuit schematics of the regenerative current mirror

When VB1=0.42V, the close-loop gain of the current mirror is ~5. The bias current in both columns are about 1uA (rather than 1uA and 5uA).

Small current pulses I1 are send into PMOSG1 net. As shown in the lower pane of Fig. 3, the current pulses are amplified by about factor of 5 in the drain currents of M13 and M14, respectively.

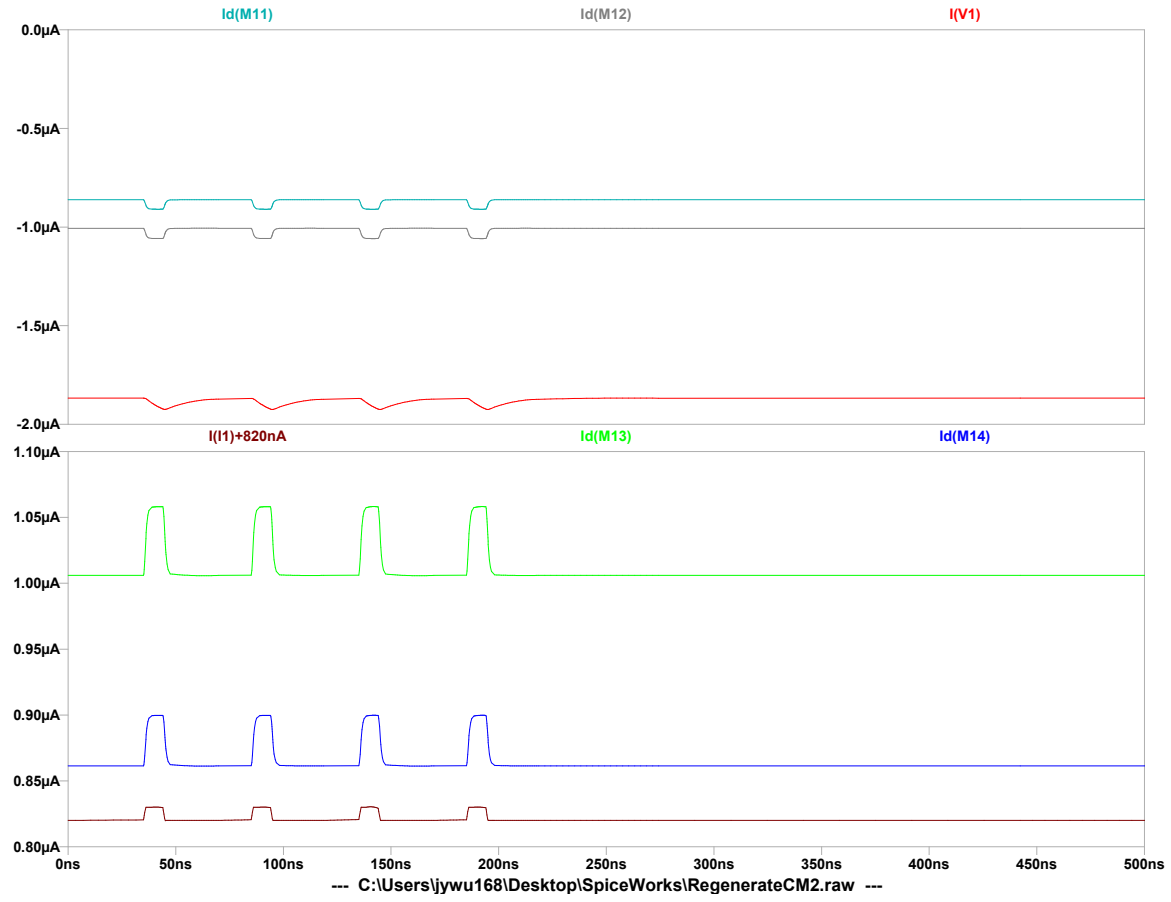

Fig. 3. Simulation results of the gain stage

The bias currents flowing through M11 and M12 are about 1µA each, while the total current flowing out the power supply V1 is about 2µA as shown in the top pane of Fig. 3. At 1.2V power supply voltage, the power consumption is around 2.4 µW for a single gain stage.

## III. An Application

For applications with very weak input signal such as silicon pixel detectors, multiple gain stages are needed. A circuit with two regenerative current mirror stages AC coupled together is shown in Fig. 4. When a small current pulse in I1 is fed into the VPG1 net, it is amplified with two gain stages as well as the AC coupling stages, and a logic level output is anticipated at net VW5.

The simulation results of the signal gain and power consumption are shown in Fig. 5

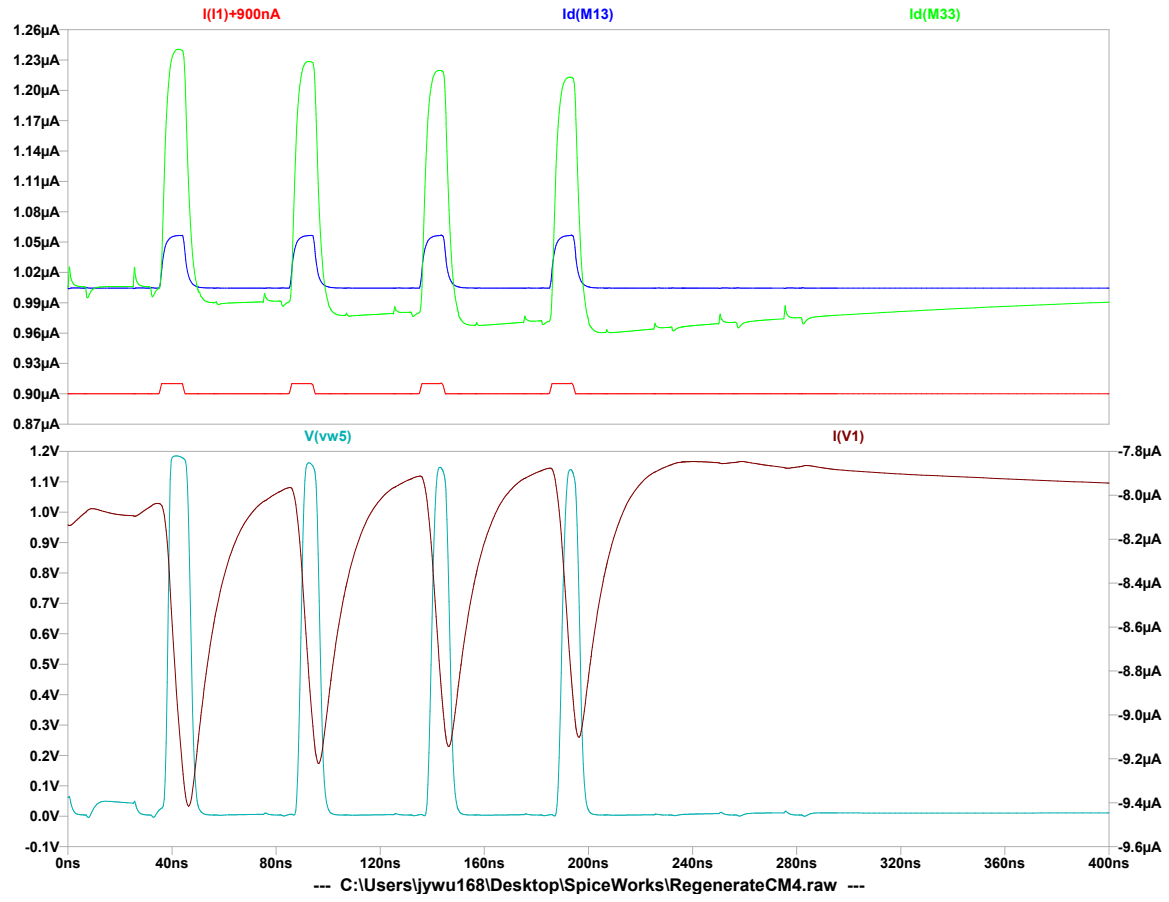

Fig. 5. Simulation results of the two-stage amplifier

The total charge of the input pulse is 10ns x 10nA = 0.1fC. (~625 e-). After two stages (with AC coupling), a logic level (1.2V) can be generated. The total power consumption is ~9.6uW (8uA x 1.2V).

## IV. Conclusions

A low power amplifier suitable for weak signals from silicon pixel detectors is design and simulated with the regenerative current mirror topology. The low power feature is realized due to two characters of this design: current amplification and positive feedback.

The amplifier can be coupled to the pseudo-thyristor a novel circuit we developed replacing conventional discriminator, (which will be described in another document). This provides a full chain solution for pixel detector with power consumption at the level of 10µW per channel.

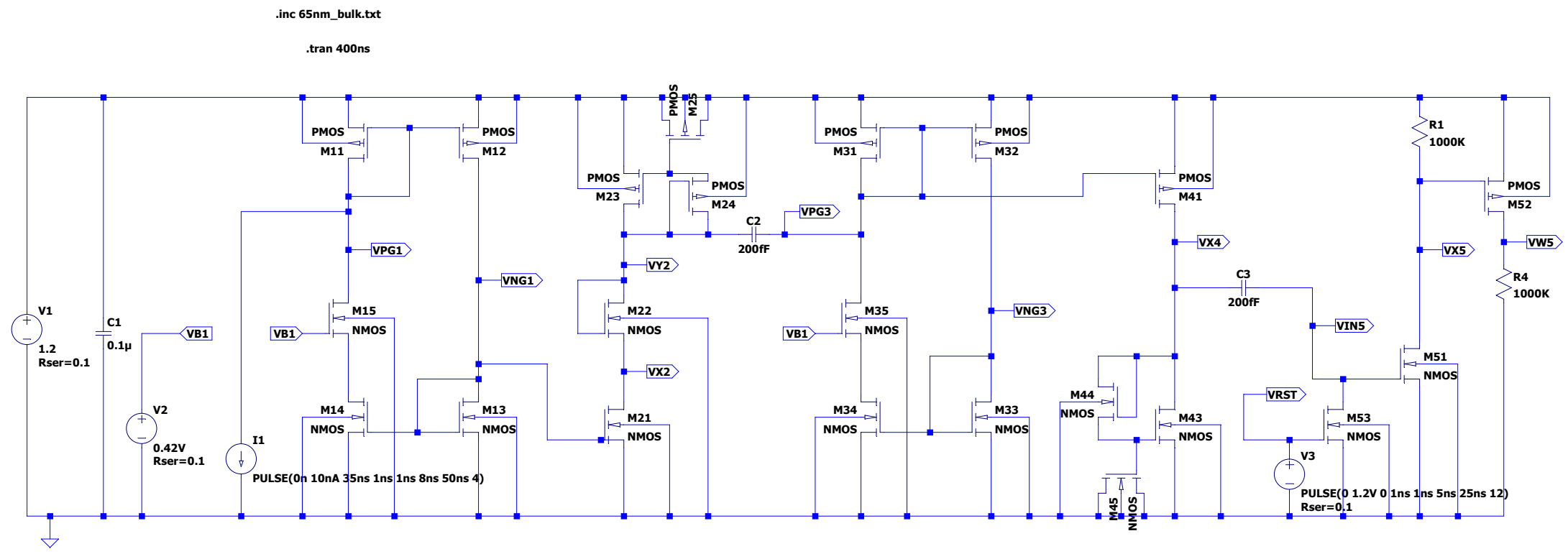

Fig. 4. Two AC coupled stages of regenerative current mirror with a logic level output buffer